\documentclass[12pt]{iopart}

\usepackage{cite}
\usepackage{iopams}
\usepackage{epsfig} 
\usepackage{subfigure}

\begin{document}

\title[The quantum $J_{1}^{XXZ}$--$J_{2}^{XXZ}$ spin-$1$ Heisenberg model on the square lattice]
{Effect of anisotropy on the ground-state magnetic ordering of the spin-one 
 quantum $J_{1}^{XXZ}$--$J_{2}^{XXZ}$ model on the square lattice}

\author{R F Bishop$^{1,2}$, P H Y Li$^{1,2}$, R Darradi$^{3}$, J Richter$^{3}$ and C E Campbell$^{2}$}
\address{$^{1}$ School of Physics and Astronomy, Schuster Building, The University of Manchester, Manchester, M13 9PL, UK}
\address{$^{2}$ School of Physics and Astronomy, University of Minnesota, 116 Church Street SE, Minneapolis, Minnesota 55455, USA}
\address{$^{3}$ Institut f\"ur Theoretische Physik, Universit\"at Magdeburg, 39016 Magdeburg, Germany}

\begin{abstract}
We study the zero-temperature phase diagram of the $J_{1}^{XXZ}$--$J_{2}^{XXZ}$ 
Heisenberg model for spin-1 particles on an infinite square lattice interacting 
via nearest-neighbour ($J_1 \equiv 1$) and next-nearest-neighbour ($J_2 > 0$) bonds.    
Both bonds have the same $XXZ$-type anisotropy in spin space.  The effects on the 
quasiclassical N\'{e}el-ordered and collinear stripe-ordered states of varying the 
anisotropy parameter $\Delta$ is investigated using the coupled cluster method carried 
out to high orders.  By contrast with the spin-$\frac{1}{2}$ case studied previously, 
we predict no intermediate disordered phase between the N\'{e}el and collinear 
stripe phases, for any value of the frustration $J_2/J_1$, for either the 
$z$-aligned ($\Delta > 1$) or $xy$-planar-aligned ($0 \leq \Delta < 1$) states.  
The quantum phase transition is determined to be first-order for all values of 
$J_2/J_1$ and $\Delta$.  The position of the phase boundary $J_{2}^{c}(\Delta)$ 
is determined accurately.  It is observed to deviate most from its classical 
position $J_2^c = \frac{1}{2}$ (for all values of $\Delta > 0$) at the Heisenberg 
isotropic point ($\Delta = 1$), where $J_{2}^{c}(1) = 0.55 \pm 0.01$.  By contrast, 
at the $XY$ isotropic point ($\Delta = 0$), we find $J_{2}^{c}(0) = 0.50 \pm 0.01$.  
In the Ising limit ($\Delta \rightarrow \infty$) $J_2^c \rightarrow 0.5$ as 
expected.
\end{abstract}


\section{Introduction}
\label{intro}
In a recent paper~\cite{Bi:2008_spinHalf_J1J2anisotrtopy} we have used the  
coupled cluster method (CCM)~\cite{Bi:1991,Bi:1998_b,Fa:2004} to study the 
influence of spin anisotropy on the ground-state (gs) magnetic ordering of an 
anisotropic version (viz., the $J_{1}^{XXZ}$--$J_{2}^{XXZ}$ model) of the 
well-known $J_{1}$--$J_{2}$ model on the infinite two-dimensional (2D) square 
lattice, described below, for particles with spin quantum number $s = \frac{1}{2}$.  
In the present paper we further the investigation of the $J_{1}^{XXZ}$--$J_{2}^{XXZ}$ 
model by replacing the spin-$\frac{1}{2}$ particles by particles with $s = 1$.

The main purpose of the previous paper was to examine carefully the role of 
spin anisotropy in tuning the quantum fluctuations that play such a key role in 
determining the quantum phase diagram of the pure (spin-isotropic) $J_{1}$--$J_{2}$ model 
that has become an archetypal model for discussing the subtle interplay between the 
effects due to quantum fluctuations and frustration, as discussed below.  While increasing 
the spin quantum number $s$ is, of course, expected to reduce the effects of 
quantum fluctuations, new and unexpected phenomena may also arise.  Thus, a well-known example 
of such  new behaviour emerging when $s$ is increased 
is the appearance of the gapped Haldane phase~\cite{Ha:1983} in 
$s = 1$ one-dimensional (1D) chains, which is not present in their $s = \frac{1}{2}$ 
counterparts.

The basic (spin-isotropic) $J_{1}$--$J_{2}$ model with nearest-neighbour (NN) 
and next-nearest-neighbour (NNN) antiferromagnetic exchange interactions, of 
strengths $J_1$ and $J_2$ respectively, has been extensively studied both 
theoretically~\cite{Ch:1988,Io:1988,Ch:1990,Schulz:1992,Ri:1993,Iv:1994,Bi:1998,Be:2002,Ro:2003,Sin:2003,Zh:2003,Ca:2004,We:2005,Sp:2005,Sir:2006} 
and experimentally~\cite{Mel:2000,Mel:2001,Ca:2002,Ka:2004}.  Many of the earlier 
studies were motivated, at least in part, by the hope of shedding light
on the possible link between antiferromagnetism and the onset of superconductivity at high temperature   
in the doped cuprate materials whose undoped precursors are seemingly well described 
by the $s = \frac{1}{2}$ version of the $J_{1}$--$J_{2}$ model on the square lattice in 
two dimensions~\cite{Ch:1990,And:1987,And:1987_PRL58,Ba:1988}.  
The recent discovery of several other quasi-2D materials that are realizations 
of the $J_{1}$--$J_{2}$ model, has only served to extend the theoretical interest in the model.  

Some of the actual magnetic compounds that can be well described by the $s=\frac{1}{2}$ 
$J_{1}$--$J_{2}$ model are La$_{2}$CuO$_{4}$~\cite{Ba:1988} for small values of
$J_{2}/J_{1}$, and  Li$_{2}$VOSiO$_{4}$ and Li$_{2}$VOGeO$_{4}$~\cite{Mel:2000,Mel:2001} for 
large values of $J_{2}/J_{1}$.  Other such materials include the compounds VOMoO$_{4}$~\cite{Ca:2002} and 
Pb$_{2}$VO(PO$_{4}$)$_{2}$~\cite{Ka:2004}.  The compound VOMoO$_{4}$ is interesting because 
its exchange couplings appear to be more than an order of magnitude larger than those of 
Li$_{2}$VOSiO$_{4}$, even though the structures of the two compounds are closely related.
Similarly, the compound Pb$_{2}$VO(PO$_{4}$)$_{2}$ also has a structure closely related to that of 
Li$_{2}$VOSiO$_{4}$, but it appears to have a ferromagnetic NN exchange coupling ($J_{1} < 0$) frustrated 
by an antiferromagnetic NNN exchange coupling ($J_{2} > 0$), with $|J_{2}/J_{1}| \approx 1.5$.  
By contrast, although all of the other compounds mentioned above are also examples of 
quasi-2D frustrated spin-$\frac{1}{2}$ magnets, they have NN and NNN exchanges that are 
both antiferromagnetic.

For the past few decades, a great deal of attention has also been devoted to magnetic 
materials with spin-1 ions,  
such as the linear chain systems including CsNiCl$_{3}$~\cite{St:1987} with a weak 
axial anisotropy, CsFeBr$_{3}$~\cite{Do:1988} with a strong planar anisotropy and 
the complex materials NENP (Ni(C$_{2}$H$_{8}$N$_{2}$)$_{2}$NO$_{2}$(ClO$_{4}$))~\cite{Ren:1988} 
with a weak planar anisotropy and 
NENC (Ni(C$_{2}$H$_{8}$N$_{2}$)$_{2}$Ni(CN$_{4}$))~\cite{Or:1995} with a strong 
planar anisotropy; as well as the 2D Heisenberg antiferromagnet 
K$_{2}$NiF$_{4}$~\cite{Bir:1970}.  The spin gaps observed in 
CsNiCl$_{3}$ and NENP are believed to be examples of the integer-spin 
gap behaviour predicted by Haldane~\cite{Ha:1983}; whereas half-odd-integer 
spin sytems are gapless.  Another new spin-gapped material is the 2D triangular 
lattice antiferromagnet NiGa$_{2}$S$_{4}$~\cite{Na:2005} which, it has been 
argued~\cite{Ts:2006,Bh:2006}, may be a ``spin nematic''~\cite{Ch:1991}.  It is 
clear, therefore, that the theoretical study of 2D spin-1 quantum magnets is 
worthy of pursuit.

In this context we note the recent discovery of superconductivity with a 
transition temperature at $T_c \approx 26\,$K in the layered iron-based compound LaOFeAs, 
when doped by partial substitution of the oxygen atoms by fluorine atoms~\cite{KWHH:2008}, 
La[O$_{1-x}$F$_x$]FeAs, with $x \approx$ 0.05--0.11.
This has been followed by the rapid discovery of superconductivity at even higher values of $T_c$ 
($\gtrsim 50\,$K) in a broad class of similar doped quaternary oxypnictide compounds.  Enormous interest has 
thereby been engendered in this class of materials.  Of particular relevance to the present 
work are the very recent first-principles calculations~\cite{MLX:2008} showing that the undoped 
parent precursor material LaOFeAs is well described by the spin-1 $J_{1}$--$J_{2}$ model on the 
square lattice with $J_1 > 0$, $J_2 > 0$, and $J_{2}/J_{1} \approx 2$.  Broadly similar 
conclusions have also been reached by other authors~\cite{SA:2008}.

Many of the above quasi-2D magnetic materials, and many others like them, display 
interesting gs phases, often with subtle quantum phase transitions between them.  
Generically, the interplay between reduced dimensionality, competing interactions and 
strong quantum fluctuations, seems to generate a number of new states of condensed matter 
with orderings that differ from the usual states of quasiclassical long-range order (LRO).  Thus, for 
high-temperature superconductivity, for example, Anderson~\cite{And:1987}  has  suggested
that quantum spin fluctuations and frustration due to doping could lead to the 
collapse of the 2D N\'{e}el-ordered antiferromagnetic phase present at zero 
doping, and that this could be a mechanism that drives the superconducting behaviour.  
This, and many similar experimental observations for other magnetic materials of 
reduced dimensionality, has intensified the study of order-disorder quantum 
phase transitions.  Thus, low-dimensional quantum antiferromagnets have attracted 
much recent attention as model systems in which strong quantum fluctuations 
might be able to destroy magnetic LRO in the ground state (GS).  
In the present paper we consider a system of $N \rightarrow \infty$ spin-1  
particles on a spatially isotropic 2D square lattice.

The isotropic Heisenberg antiferromagnet with only nearest-neighbour (NN) bonds, 
all of equal strength ($J_{1} > 0$), exhibits magnetic LRO at zero temperature on such 
bipartite lattices as the square lattice considered here.  A key mechanism 
that can then serve to destroy the LRO for such systems (with a given lattice and spins of 
a given spin quantum number $s$) is the introduction of competing or 
frustrating bonds on top of the NN bonds.  The interested reader is 
referred to~\cite{Scholl:2004,Ma:1991} for a more detailed 
discussion of 2D spin systems in general.

In this context, and as we have already noted above, 
an archetypal frustrated model of the above type that has attracted much theoretical 
attention in recent years 
is the 2D $J_{1}$--$J_{2}$ model on a square lattice with both 
NN and NNN antiferromagnetic interactions, with 
strength $J_{1} > 0$ and $J_{2} > 0$ respectively.  The NN bonds $J_{1} > 0$ 
promote N\'{e}el antiferromagnetic order, while the NNN bonds $J_{2} > 0$ act to frustrate 
or compete with this order.  All such frustrated quantum magnets continue to be 
of great theoretical interest because of the possible spin-liquid and other such 
novel magnetically disordered phases that they can exhibit (and see, e.g.,~\cite{RiScHo:2004,Sach:2004,Mis:2005}).  

The properties of the $s=1/2$ $J_{1}$--$J_{2}$ model on the 2D square lattice are 
well understood in the limits when $J_{2}=0$ or $J_{1}=0$.  For the case when $J_{2}=0$, 
and the classical GS is perfectly N\'{e}el-ordered, the quantum fluctuations 
are not sufficiently strong enough to destroy the N\'{e}el LRO, although the staggered magnetization 
is reduced to about 61\% of its classical value.  The opposite limit of large $J_{2}$ is a classic 
example~\cite{Ch:1990} of the phenomenon of order by disorder~\cite{Vi:1977,Shen:1982}.    
Thus, in the case where $J_{1} \rightarrow 0$ with $J_{2} \neq 0$ and fixed, the 
two sublattices each order antiferromagnetically at the classical level, but in 
directions which are independent of each other.  This degeneracy is lifted by quantum 
fluctuations and the GS becomes magnetically ordered collinearly as a stripe phase 
consisting of successive alternating rows (or columns) of parallel spins.
It is by now also widely accepted 
that the $s=1/2$ $J_{1}$--$J_{2}$ model exhibits the above two quasiclassical 
antiferromagnetic phases with LRO at small and at large $J_{2}$ separated by an intermediate 
quantum paramagnetic phase without magnetic LRO in the parameter region 
$J_{2}^{c_{1}} < J_{2} < J_{2}^{c_{2}}$ where $J_{2}^{c_{1}} \approx 0.4J_{1}$ 
and $J_{2}^{c_{2}} \approx 0.6J_{1}$.  The GS at low $J_{2} < J_{2}^{c_{1}}$ 
exhibits N\'{e}el-ordered magnetic LRO (with a wave vector $Q = (\pi,\pi)$), 
whereas the GS at large $J_{2} > J_{2}^{c_{2}}$ exhibits collinear stripe-ordered 
magnetic LRO (with a wave vector $Q = (\pi,0)$) or $Q = (0,\pi)$).

Given the key role played by quantum fluctuations in determining the gs structure of frustrated 
magnets, it is clearly of central interest to focus special attention on the 
various means by which we may vary or ``tune'' them.  Clearly, as we have already noted, an 
increase in the spin quantum number $s$ is expected to decrease their strength.  Thus, for example, 
for the simple case of the isotropic Heisenberg model on the square lattice with NN 
bonds all of the same strength, whereas the quantum fluctuations reduce the perfect N\'{e}el 
ordering in the classical case (i.e., $s \rightarrow \infty$) so that the staggered 
magnetization is only about 61\% of its classical value for the $s = \frac{1}{2}$ case as noted above, 
the corresponding reduction in the $s = 1$ case is less, namely to about 80\% of the classical 
value (and see~\cite{Fa:2001} and references cited therein).  One of the goals of the 
present paper is to investigate similarly the effect of increasing $s$ for the 
archetypal $J_{1}$--$J_{2}$ model on the 2D square lattice.  In order to do so it 
is convenient to consider at the same time any other means to ``tune'' the quantum 
fluctuations.  In particular, we note that besides changing $s$  
or the dimensionality and lattice type of the system, and apart from varying the 
relative strengths of the competing exchange interactions, another key mechanism to tune 
the quantum fluctuations is the introduction of anisotropy, either in 
real space~\cite{Ne:2003,Si:2004,Star:2004,Mo:2006,Bi:2007_j1j2j3_spinHalf,Bi:2008} 
or in spin space~\cite{Be:1998,Ro:2004,Via:2007,Da:2004}, into the existing exchange bonds.  

Turning first to the case of anisotropy in real (crystal lattice) space, we note that 
Nersesyan and Tsvelik~\cite{Ne:2003} have recently introduced and studied an interesting generalization 
of the pure $J_{1}$--$J_{2}$ model for the $s = \frac{1}{2}$ case in order to investigate the 
effects of spatial anisotropy on the quantum fluctuations in the model.  
This extended model, the so-called $J_{1}$--$J_{1}'$--$J_{2}$ model, has been further 
studied by other groups for both the $s = \frac{1}{2}$~\cite{Si:2004,Star:2004,Mo:2006,Bi:2007_j1j2j3_spinHalf} 
and the $s=1$~\cite{Bi:2008} cases.  This generalization of the 2D $J_{1}$--$J_{2}$ model 
introduces a spatial anisotropy on the 
square lattice by allowing the NN bonds to have different strengths $J_{1}$ 
and $J_{1}'$ in the two orthogonal spatial lattice dimensions, while keeping all 
of the NNN bonds across the diagonals to have the same strength $J_{2}$.  In previous 
work of our own~\cite{Bi:2007_j1j2j3_spinHalf,Bi:2008} on this 
$J_{1}$--$J_{1}'$--$J_{2}$ model we studied the effect of the coupling $J_{1}'$ on 
the semiclassical N\'{e}el-ordered and stripe-ordered phases.  For the $s = \frac{1}{2}$
case, we found that the quantum critical points for both of these phases with 
LRO increase as the coupling ratio $J_{1}'/J_{1}$ is increased, and an 
intermediate phase with no magnetic LRO only emerges when $J_{1}'/J_{1} \gtrsim 0.6$, 
with strong indications of a quantum triple point at  
$J_{1}'/J_{1} \approx 0.60, J_{2}/J_{1} \approx 0.33$.  For $J_{1}'/J_{1}=1$, 
the results agree with the previously known results of the $J_{1}$--$J_{2}$ 
model described above.  

By contrast, for the $s=1$ case, we found no evidence for an 
intermediate phase between the N\'{e}el and stripe states, as compared with 
all previous results for the corresponding $s = \frac{1}{2}$ case.  However, for 
the $s = 1$ case we found instead strong evidence for a quantum tricritical point 
at $J_{1}'/J_{1} \approx 0.66, J_{2}/J_{1} \approx 0.35$, where  
a line of second-order phase transitions between the quasiclassical N\'{e}el-ordered  
and stripe-ordered phase (for $J_{1}'/J_{1} \lesssim 0.66$) meets a line of first-order 
phase transitions between the same two states (for $J_{1}'/J_{1} \gtrsim 0.66$).  
For $J_{1}'/J_{1} = 1$ the results obviously reproduce those of the usual 
spin-1 $J_{1}$--$J_{2}$ model, for which $J_{2}^{c}/J_{1} \approx 0.55 \pm 0.01$.

Finally, we turn to the main subject of interest in this paper, namely to further the 
study of the 2D spin-1 $J_{1}$--$J_{2}$ model on the square lattice by introducing 
anisotropy in spin space.  While the influence of the spin anisotropy on 
the $s = \frac{1}{2}$ $J_{1}$--$J_{2}$ model on the square lattice 
has been studied by various groups~\cite{Be:1998,Da:2004,Ro:2004,Via:2007}, 
including ourselves~\cite{Bi:2008_spinHalf_J1J2anisotrtopy}, relatively little is 
known for the $s=1$ case.

Our aim here is to further the study of the $J_{1}^{XXZ}$--$J_{2}^{XXZ}$ model 
for the $s=1$ case, by making use of the coupled cluster method (CCM) carried out to high 
orders by making use of supercomputing resources.  The 
CCM (see~\cite{Bi:1991,Bi:1998_b,Fa:2004} and references cited therein) is 
one of the most powerful and most universally applicable of all known {\it ab initio\/} techniques of 
modern microscopic quantum many-body theory. It is also one of the most accurate 
methods available at  attainable levels of computational implementation.  We note, in the 
present context, that the CCM is a particularly effective tool for studying highly frustrated 
quantum magnets, where such other numerical methods as the quantum Monte Carlo method and the 
exact diagonalization method are often severely limited in practice, e.g., by the 
``minus-sign problem'' for the former case, and the very small sizes of the spin systems that can be 
handled in practice with available computing resources for the latter.   This is especially 
true for spin systems with spin quantum number $s > \frac{1}{2}$, as are of interest here.  
The CCM has been applied successfully on many previous occasions to calculate the ground-state 
and excited-state properties of a diverse array of quantum spin 
systems~\cite{Bi:2008_spinHalf_J1J2anisotrtopy,Fa:2004,Bi:1998,Fa:2001,Bi:2007_j1j2j3_spinHalf,
Bi:2008,Da:2004,RoHe:1990,BPX:1991,BiHX:1994,Xi:1994,Burs:1995,
Ze:1998,RoLiBi:1999,Kr:2000,Fa:2002,IvRJ:2002,Da:2005,Schm:2006,RDZB:2007,Zi:2008}.

\section{The model}
Exactly as for the $s=\frac{1}{2}$ case that we studied earlier~\cite{Bi:2008_spinHalf_J1J2anisotrtopy}, the 
$s=1$ $J_{1}$--$J_{2}$ Heisenberg model employed here has two kinds of exchange 
bonds, namely the NN $J_{1}$ bonds along both the row and the column directions of the square lattice, 
and the NNN $J_{2}$ bonds along the diagonals of the squares.  The model is then generalized by including 
an anisotropy in spin space in both types of bonds.  The anisotropy parameter $\Delta$ 
is assumed to be the same in both exchange terms, thus producing the so-called 
$J_{1}^{XXZ}$--$J_{2}^{XXZ}$ model, with a Hamiltonian given by 
\begin{eqnarray}
H =& & J_{1}\sum_{\langle i,j \rangle}(s^{x}_{i}s^{x}_{j}+s^{y}_{i}s^{y}_{j}+\Delta s^{z}_{i}s^{z}_{j}) \nonumber\\
   & + & J_{2}\sum_{\langle\langle i,k \rangle\rangle}(s^{x}_{i}s^{x}_{k}+s^{y}_{i}s^{y}_{k}+\Delta s^{z}_{i}s^{z}_{k})\,,  \label{H}
\end{eqnarray}
where the sums over $\langle i,j \rangle$ and $\langle\langle i,k \rangle\rangle$ 
run over all NN and NNN pairs respectively, counting each bond once and once only. 
Both exchange couplings are assumed 
to be antiferromagnetic here (i.e., $J_{1}>0$ and $J_{2}>0$), and henceforth the energy scale is 
set by putting $J_{1}=1$.  We shall also only be concerned here with the case $\Delta \geq 0$.  

The model has two types of classical ground state (GS), namely a $z$-aligned 
state for $\Delta > 1$ and an $xy$-planar-aligned state for 
$0 < \Delta < 1$.  Since all directions in the $xy$-plane in spin space are equivalent, 
we may choose the direction arbitrarily for the $xy$-planar-aligned state to be the $x$-direction, say.  
Both of these $z$-aligned and $x$-aligned ground states further divide into a 
N\'{e}el ($\pi,\pi$) state and collinear stripe states (columnar stripe
($\pi,0$) and row stripe ($0,\pi$)).  There is clearly a symmetry
under the interchange of rows and columns, and hence we only consider the columnar 
stripe state.  The N\'{e}el states are the classical GS for 
$J_{2} < \frac{1}{2}J_{1}$, and the collinear stripe states are the classical 
GS for $J_{2} > \frac{1}{2}J_{1}$.  The (first-order) classical phase transition between 
these states of perfect classical LRO occurs precisely at $J^{c}_{2}=\frac{1}{2} J_{1}$, .

\section{The coupled cluster method}
\label{CCM}
We now briefly describe the CCM formalism.  For further details interested readers are referred, for 
example, to~\cite{Bi:1991,Bi:1998_b,Fa:2004} and references cited therein.  

In order to use the CCM the first step is always the choice of  a normalized model (or reference) 
state $|\Phi\rangle$ which is required to act as a cyclic vector (or, more physically, as a 
generalized vacuum state) with respect to a complete set of mutually commuting multi-configurational 
creation operators, $C^{+}_{I} \equiv (C^{-}_{I})^{\dagger}$ that need to be chosen simultaneously.  
The index $I$ here is a set-index that gives a complete labelling of the many-particle configuration 
created in the state $C^{+}_{I}|\Phi\rangle$.  The requirements on $\{|\Phi\rangle; C^{+}_{I}\}$ 
are that any many-particle state can be exactly decomposed as a unique linear combination 
of the states $\{C^{+}_{I}|\Phi\rangle\}$, together with the conditions,
\begin{equation}
\langle\Phi| C^{+}_{I} = 0 = C^{-}_{I}|\Phi\rangle \quad \forall \emph{I} \neq 0\,; \quad C^{+}_{0} \equiv 1 \, ,
\end{equation}
\begin{equation}
[C^{+}_{I},C^{+}_{J}] = 0 = [C^{-}_{I},C^{-}_{J}]\,.  \label{C_comm}
\end{equation}

The exact many-body gs ket and bra states, whose solutions we seek via the CCM calculation at hand, 
satisfy the respective Schr\"{o}dinger equations,
\numparts
\begin{eqnarray}
H|\Psi\rangle &= E|\Psi\rangle \,,    \label{Sch_eq_ket} \\      
\langle\tilde{\Psi}|H &=E\langle\tilde{\Psi}|\,  \label{Sch_eq_bra}\,,
\end{eqnarray}
\endnumparts
respectively, with the normalization defined by $\langle\tilde{\Psi}|\Psi\rangle = 1$ 
[i.e., with $\langle\tilde{\Psi}|=(\langle\Psi|\Psi\rangle)^{-1}\langle\Psi|$], 
and with $|\Psi\rangle $ itself satisfying the intermediate normalization 
condition $\langle\Phi|\Psi\rangle=1=\langle\Phi|\Phi\rangle$.  In terms 
of the set $\{|\Phi\rangle; C^{+}_{I}\}$, the CCM now employs an exponential 
parametrization for the exact gs ket energy eigenstate, 
\numparts
\begin{eqnarray}
|\Psi\rangle &= \mbox{e}^{S}|\Phi\rangle\,, \quad S & = \sum_{I\neq0}{\cal S}_{I}C^{+}_{I}\,,  \label{ccm_para_ket}
\end{eqnarray}
that lies at the heart of the method.  Its counterpart for the exact gs bra energy eigenstate is chosen as
\begin{eqnarray}
\langle\tilde{\Psi}|=\langle\Phi|\tilde{S}\mbox{e}^{S}\,, \quad  \tilde{S} & = 1 + \sum_{I\neq0}\tilde{{\cal S}_{I}}C^{-}_{I}\,.    \label{ccm_para_bra}
\end{eqnarray}
\endnumparts

The gs CCM correlation operators, $S$ and $\tilde{S}$, contain 
the real c-number correlation coefficients, ${\cal S}_{I}$ and $\tilde{{\cal S}_{I}}$,  
that need to be calculated.  Clearly, once they are known, all other gs properties 
of the many-body system can be derived from them.  In order to find them 
we simply insert the parametrizations (\ref{ccm_para_ket}) and (\ref{ccm_para_bra}) into 
the Schr\"{o}dinger equations (\ref{Sch_eq_ket}) and (\ref{Sch_eq_bra}), and then project onto the complete 
sets of states $\langle\Phi|C^{-}_{I}$ and $C^{+}_{I}|\Phi\rangle$, respectively.  
Completely equivalently, we may simply demand that the gs energy expectation 
value, $\bar{H} \equiv \langle\tilde{\Psi}|H|\Psi\rangle$, is minimized with respect 
to the entire set $\{{\cal S}_{I}, \tilde{{\cal S}_{I}}\}$.  In either case 
we are easily led to the equations
\numparts
\begin{eqnarray}
\langle \Phi|C^{-}_{I}\mbox{e}^{-S}H\mbox{e}^{S}|\Phi\rangle & = 0\;; \quad  \forall I \neq 0\,,   \label{ket_coeff} \\
\langle\Phi|\tilde{S}\mbox{e}^{-S}[H, C^{+}_{I}]\mbox{e}^{S}|\Phi\rangle & = 0\;; \quad \forall I \neq 0 \,,  \label{bra_coeff}
\end{eqnarray}
\endnumparts
which are first derived using computer algebra and then 
solved for the set $\{{\cal S}_{I}, \tilde{{\cal S}_{I}}\}$
within specific truncation schemes described below, 
by making use of parallel computing routines~\cite{ccm}.  
Equation (\ref{ket_coeff}) also shows that the gs energy at the stationary point has 
the simple form
\begin{equation}
E = E(\{{\cal S}_{I}\})=\langle\Phi|\mbox{e}^{-S}H\mbox{e}^{S}|\Phi\rangle\,.   \label{Energy}
\end{equation}
It is important to realize that this bi-variational formulation does not 
necessarily lead to an upper bound for $E$ when the summations for $S$ and 
$\tilde{S}$ in (\ref{ccm_para_ket},b) are truncated, due to the lack of 
manifest Hermiticity when such approximations are made.  Nonetheless, 
one can prove~\cite{Bi:1998_b} that the important Hellmann-Feynman 
theorem {\it is\/} preserved in all such approximations.

Equation (\ref{ket_coeff}) represents a coupled set of nonlinear 
multinomial equations for the c-number correlation coefficients $\{ {\cal S}_{I} \}$.  
The nested commutator expansion of the similarity-transformed Hamiltonian,
\begin{equation}
\mbox{e}^{-S}H\mbox{e}^{S} = H + [H,S] + \frac{1}{2!}[[H,S],S] + \cdots \,,  \label{H_Sim_xform}
\end{equation}
and the fact that all of the individual components of $S$ in the expansion of 
(\ref{ccm_para_ket}) commute with one another by construction, as in (\ref{C_comm}), 
together imply that each element of $S$ in (\ref{ccm_para_ket}) is linked directly 
to the Hamiltonian in each of the terms in (\ref{H_Sim_xform}).  Each of the 
coupled equations (\ref{ket_coeff}) is hence of Goldstone {\it linked-cluster\/} type,   
which thereby guarantees that all extensive variables, such as the energy, 
scale linearly with particle number, $N$.  Thus, at any level of approximation 
obtained by truncation in the summations on the index $I$  in (\ref{ccm_para_ket}) and (\ref{ccm_para_bra}), 
we may always work safely from the outset in the limit $N \rightarrow \infty$ 
of an infinite system, as we do in all our calculations below.  It is also important to 
note that each of the linked-cluster equations (\ref{ket_coeff}) is actually of finite 
length when expanded, since the otherwise infinite series of (\ref{H_Sim_xform}) 
will always terminate at a finite order, provided only that each term in the Hamiltonian, $H$, 
contains a finite number of single-particle destruction operators defined with 
respect to the reference (vacuum) state $|\Phi\rangle$, as in the case of our Hamiltonian (\ref{H}).  

We turn now to the implementation of the CCM for quantum spin systems, for which it 
is usually convenient to take the classical ground states as 
our (initial) choices for the model state $|\Phi\rangle$.  Hence, 
we may choose here either a N\'{e}el state or a collinear (columnar) 
stripe state for $|\Phi \rangle$.  Each of these can be further sub-divided into 
a $z$-aligned choice or an $xy$-planar (say, $x$-aligned) choice, which we expect to be 
appropriate for the regions $\Delta \geq 1$ and $0 \leq \Delta \leq 1$ respectively on purely classical 
grounds.  We present results in section~\ref{results} based on all four of these 
classical ground states as choices for $|\Phi\rangle$.  In order to implement the CCM 
computationally it is very convenient to treat the spins on every lattice site in any 
chosen model state $|\Phi\rangle$ as equivalent.  In order to do so we introduce a 
different local quantization axis and a correspondingly different set of 
spin coordinates on each site, so that all spins, whatever 
their original orientations in $|\Phi\rangle$ in the global spin-coordinate system, 
align along the negative $z$-direction, say, in these local spin coordinates.  
This can always be done by defining a suitable rotation in spin space of the 
global spin coordinates at each lattice site.  Such rotations are canonical 
transformations that leave the spin commutation relations unchanged.  In these 
local spin axes where the configuration indices $I$ simply become a set of lattice 
site indices, $I \rightarrow \{k_{1},k_{2},\cdots k_{m}\}$, the generalized 
multi-configurational creation operators $C^{+}_{I}$ are simple products of 
single spin-raising operators, 
$C^{+}_{I} \rightarrow s^{+}_{k_{1}} s^{+}_{k_{2}} \cdots s^{+}_{k_{m}}$, 
where $s^{\pm}_{k} \equiv s^{x}_{k} \pm is^{y}_{k}$\,, 
and $(s^{x}_{k}, s^{y}_{k}, s^{z}_{k})$ are the usual SU(2) spin operators 
on lattice site $k$.  For the quasiclassical 
magnetically-ordered states that we calculate here, the order parameter is the 
sublattice magnetization, $M$, which is given within our local spin coordinates 
defined above as
\begin{equation}
M \equiv -\frac{1}{N} \langle\tilde{\Psi}|\sum_{k=1}^{N}s^{z}_{k}|\Psi\rangle\,.
\end{equation}

The CCM formalism is clearly exact if one includes all spin configurations $I$ in the 
expansions (\ref{ccm_para_ket}) and (\ref{ccm_para_bra}) of the $S$ and $\tilde{S}$ operators respectively.  
However, truncations are necessary in practice.  Based on a great deal of previous 
experience, we usually employ the so-called LSUB$n$ approximation scheme 
for $s=1/2$ quantum spin systems (see~\cite{Bi:2007_j1j2j3_spinHalf} and 
references cited therein), and its so-called SUB$n$--$m$ counterpart for $s=1$ systems 
(see~\cite{Bi:2008} and references cited therein).  The LSUB$n$ scheme is defined such 
that all possible multi-spin-flip correlations over different locales on the 
lattice defined by $n$ or fewer contiguous lattice sites are retained at the $n$th 
level of approximation.  For the case of spins with $s = \frac {1}{2}$, the  
multi-configurational creation operators, $C^{+}_{I}$ can contain no more than one 
spin-raising operator $s^{+}_{j}$ for each lattice site $j$.  However, the number of 
fundamental LSUB$n$ configurations for $s = 1$ becomes appreciably higher than for 
$s = \frac{1}{2}$, since each spin on each site $j$ can now be flipped twice by the spin-raising 
operators, so that in this case the  multi-configurational creation operators, $C^{+}_{I}$ 
can contain up to two spin-raising operator $s^{+}_{j}$ for each lattice site $j$.  
Thus, for systems with $s > \frac{1}{2}$ it is more practical to use the SUB$n$--$m$ scheme, in which 
all correlations involving no more than $n$ spin flips spanning a range of no more than $m$ 
adjacent lattice sites are retained.  Clearly, for spins with $s=1$, the SUB$2n$--$n$ scheme 
is fully equivalent to the LSUB$n$ scheme.  More generally for spins with arbitrary spin quantum number $s$, 
SUB$2sn$--$n$ $\equiv$ LSUB$n$.  In order to keep the number of fundamental configurations 
from growing too quickly with increasing level of approximation we set $m = n$, and thus we have the 
SUB$n$--$n$ scheme.  The approximation clearly becomes exact as $n \rightarrow \infty$.

We note that, in general terms, both the LSUB$n$ and SUB$n$--$m$ truncation schemes are 
systematic {\it localized\/} approximation hierarchies in which the truncation indices are 
physically related to the size of the clusters of spins on the lattice for which the multi-spin 
correlations are explicitly included.  Their physical motivation (and eventual justification) 
thus stems ultimately from the localized short-range nature of the underlying Hamiltonian (which, 
in the present case, involves just two-spin interactions at NN and NNN distances apart only).  The 
maximum number of spins correlated in such clusters is $n$ in both cases.  By contrast, the SUB$n$ 
scheme (which is formally equivalent to the SUB$n$--$m$ scheme in the limit $m \rightarrow \infty$) 
explicitly correlates all clusters of spins involving no more than $n$ spin-flips, regardless 
of the spatial separations of the spins within the correlated clusters.  It is important to note  
however that in {\it all\/} CCM approximations (including the LSUB$n$ and SUB$n$--$m$ schemes) 
each correlated cluster configuration retained within the correlation operator $S$ of 
(\ref{ccm_para_ket}) is actually counted an {\it arbitrarily large\/} number of times due to the 
exponentiated form in which the operator $S$ appears in the parametrization (\ref{ccm_para_ket}). 
It is precisely the exponential form that guarantees the proper counting of arbitrary multiples, 
at different positions on the lattice, of each configuration (and all products of such multiples for 
different configurations) retained in $S$, considered as {\it independent\/} excitations.  Thus, 
even though, for example, the LSUB$n$ and SUB$n$--$m$ truncation schemes are motivated by the 
inclusion of the explicit correlations within localized clusters of spins only up to a given size, 
every approximation includes configurations in which an {\it arbitrary\/} number of spins (up to all 
$N \rightarrow \infty$ spins) are correlated, albeit as (properly counted) products of independent 
sub-clusters up to a given finite size.

Table~\ref{FundConf_spin1} 
\begin{table}[t]
\begin{center}
\caption{Numbers of fundamental configurations ($\sharp$ f.c.) retained in the 
CCM SUB$n$--$n$ approximation for the $z$-aligned states and the planar $x$-aligned 
states of the $s=1$ $J_{1}^{XXZ}$--$J_{2}^{XXZ}$ model on the square lattice.}
\begin{tabular}{cccccc} \hline\hline
 & \multicolumn{2}{c}{$z$-aligned states}  &  & \multicolumn{2}{c}{planar $x$-aligned states} \\ 
\\[-7pt] \cline{1-3} \cline{5-6} \\[-6pt]
 Scheme  & \multicolumn{2}{c}{$\sharp$ f.c.}  &   &\multicolumn{2}{c}{$\sharp$ f.c.} \\
\\[-7pt] \cline{2-3} \cline{5-6} \\[-6pt]
& N\'{e}el  & stripe   &  & N\'{e}el & stripe \\ 
\\[-7pt] \cline{1-3}  \cline{4-6} \\[-6pt]
SUB$2$--$2$ & 1 & 1 &  & 2  & 3 \\
SUB$4$--$4$ & 15 & 21 & & 31 & 57 \\
SUB$6$--$6$ & 375 & 585 &  & 1085 & 2131 \\
SUB$8$--$8$ & 17864 & 29411 &  & 61904 & 123471 \\ \hline\hline
\end{tabular}
\label{FundConf_spin1}
\end{center}
\end{table}
shows the number of fundamental SUB$n$--$n$ configurations for the $z$-aligned and 
planar $x$-aligned states in the N\'{e}el and striped phases.  We see 
that the number of fundamental configurations for the planar model state at the 
SUB$8$--$8$ level of approximation is 61904 for the N\'{e}el phase and 123471 for 
the stripe phase.  The intensive calculations required at even this very high order of 
approximation are easily practicable with relatively modest supercomputing resources.  
Thus, for example, we employed 200 processors simultaneously to execute the SUB$8$--$8$ 
calculations using the planar $x$-aligned collinear stripe state as model state, and with 
this number of processors it took about six hours to solve the CCM equations 
(\ref{ket_coeff}) and (\ref{bra_coeff}) at this level of approximation for each value of 
the anisotropy parameter $\Delta$ in the Hamiltonian (\ref{H}).

Clearly, the last step in our calculations is to extrapolate the approximate SUB$n$--$n$ 
results to the exact, $n \rightarrow \infty$, limit.  We use here for the extrapolations 
of the raw SUB$n$--$n$ data the same well-tested scaling laws as we used previously in our studies of the  
$J_{1}$--$J_{1}'$--$J_{2}$ model for both the $s = \frac{1}{2}$ case~\cite{Bi:2007_j1j2j3_spinHalf} 
and the $s = 1$ case~\cite{Bi:2008}, as well as for the $s = \frac{1}{2}$ version of 
the present model~\cite{Bi:2008_spinHalf_J1J2anisotrtopy}.  
Thus, the scaling law used for the gs energy per spin, $E/N$, is 
\begin{equation}
E/N=a_{0}+a_{1}n^{-2}+a_{2}n^{-4}\;,   \label{E_scaling}
\end{equation} 
and that for the staggered magnetization, $M$, is
\begin{equation}
M=b_{0}+n^{-0.5}\left(b_{1}+b_{2}n^{-1}\right)\;.   \label{M_scaling}
\end{equation} 

In order to have a robust and stable fit to any fitting formula that contains $m$ unknown 
parameters, it is well known that it is desirable to have at least ($m+1$) data points 
(the so-called $m+1$ rule).  Both of our scaling laws (\ref{E_scaling}) and (\ref{M_scaling}) 
contain $m=3$ unknown parameters to be determined, and in all cases we have SUB$n$--$n$ data 
sets with $n=\{2,4,6,8\}$.  In all our results presented below the SUB$n$--$n$ results are 
extrapolated to the limit $n \rightarrow \infty$ using the sets with $n=\{2,4,6,8\}$ for 
both the $z$-aligned and planar $x$-aligned states.  However, we have also extrapolated $E/N$ 
and $M$ using the sets $n=\{4,6,8\}$ and $n=\{2,4,6\}$.  In all cases they lead to very 
similar results, thereby adding credence to their validity and stability.
We also note that for the corresponding $s=1/2$ model we could perform 
LSUB$n\equiv$ SUB$n$--$n$ approximation calculations for $n=\{2,4,6,8,10\}$.  
This enabled us to perform extrapolations using the sets 
$n=\{2,4,6,8\}$ and $n=\{2,4,6,8,10\}$ as well as the preferred set 
$n=\{4,6,8,10\}$.  Gratifyingly, all sets yielded very similar extrapolated 
results, even near phase boundaries and the quantum triple point, which gives us 
great confidence in the accuracy and robustness of our extrapolation scheme.

\section{Results}
\label{results}
\begin{figure*}[tbp]
\mbox{
 \subfigure[$z$-aligned states]{\label{E_Zaligned_spin1}\scalebox{0.3}{\epsfig{file=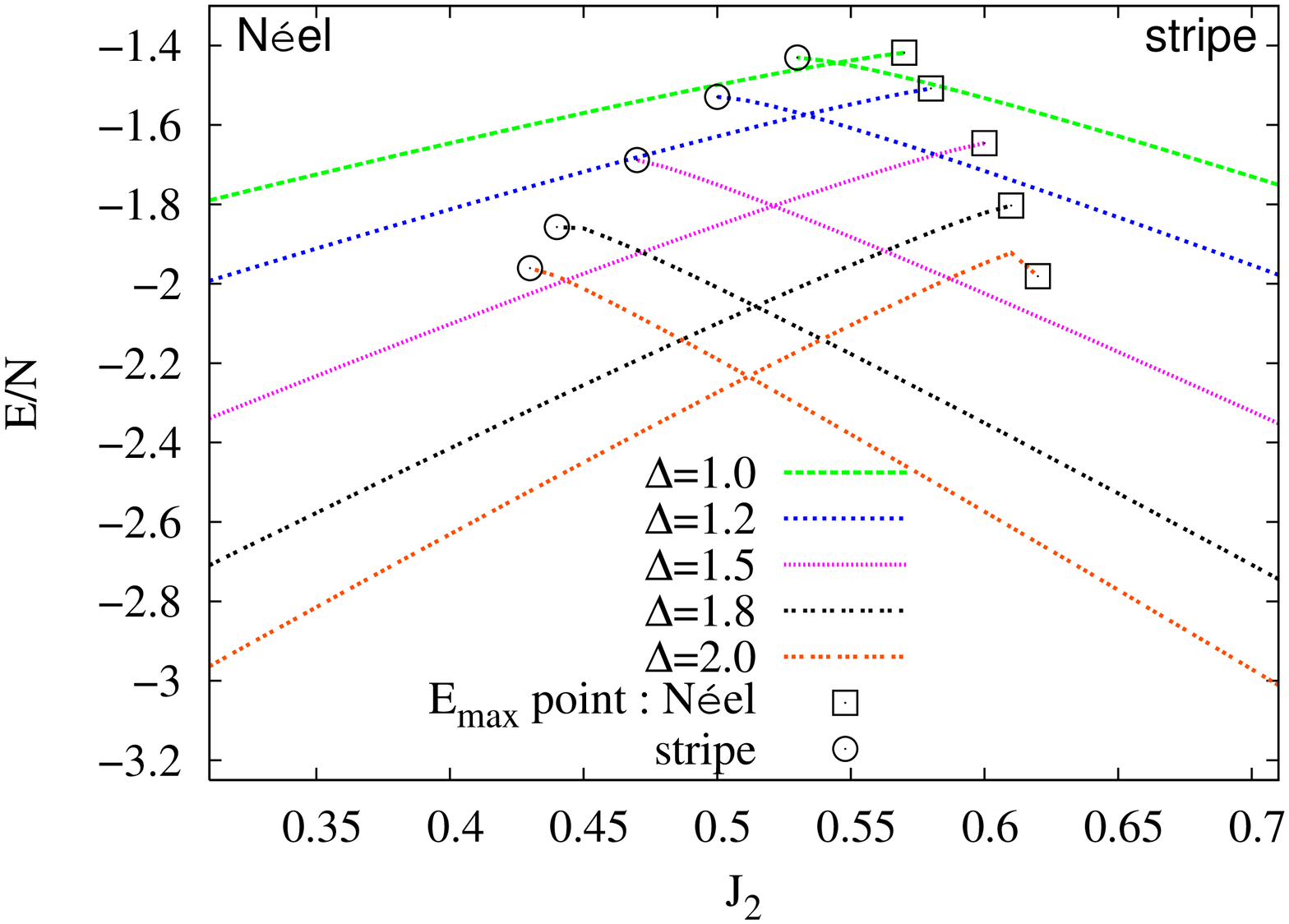,angle=270}}}
  \subfigure[planar $x$-aligned states]{\label{E_planar_spin1}\scalebox{0.3}{\epsfig{file=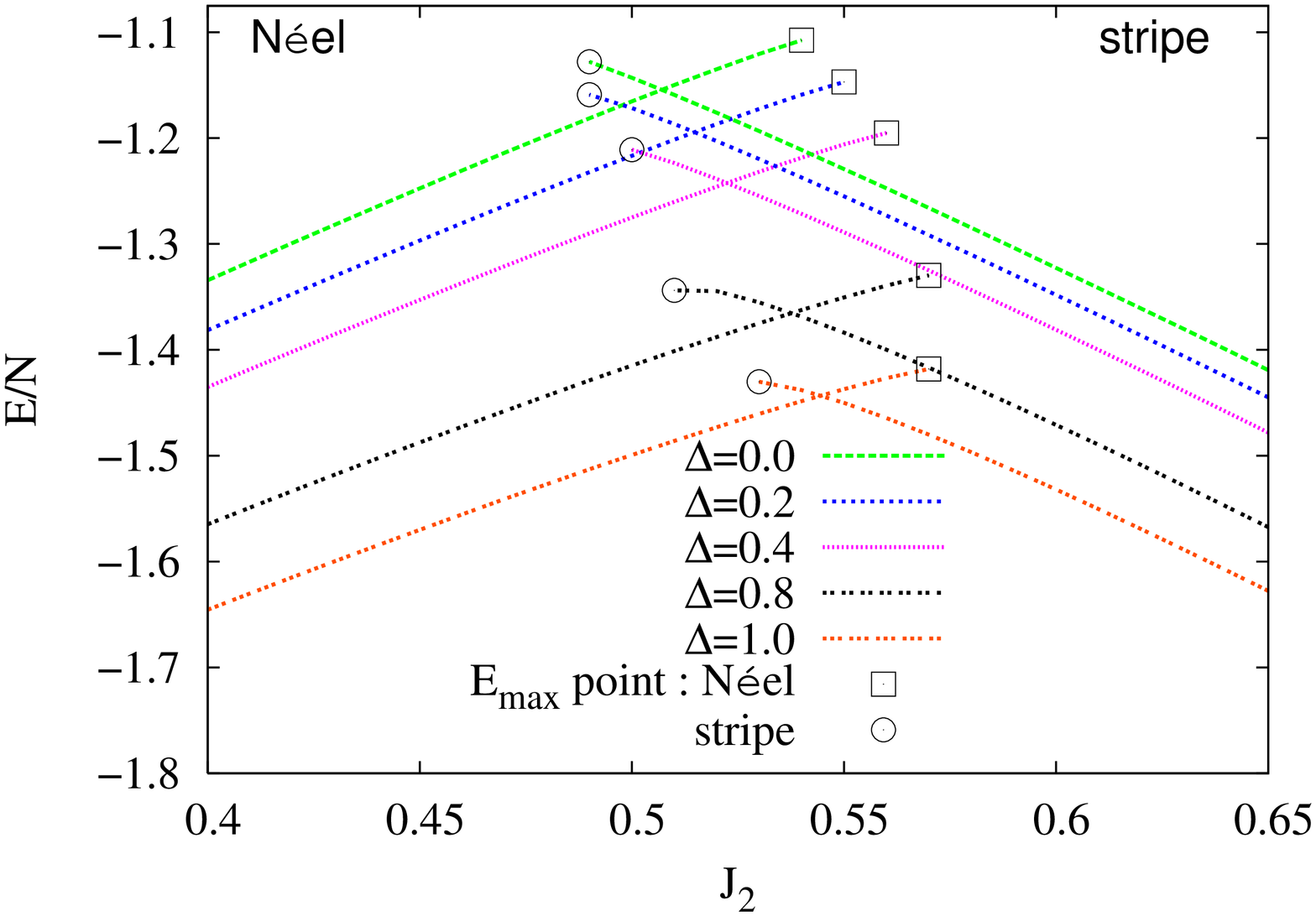,angle=270}}}
}
\caption{(Colour online) Extrapolated CCM SUB$n$--$n$ results using the $z$-aligned and 
planar $x$-aligned states for the gs energy, E/N, for the N\'{e}el and 
stripe phases of the $s=1$ $J_{1}^{XXZ}$--$J_{2}^{XXZ}$ model.  The SUB$n$--$n$ 
results are extrapolated to the limit $n \rightarrow \infty$ using the sets 
$n=\{2,4,6,8\}$ for both the $z$-aligned and planar $x$-aligned states.  
The NN exchange coupling $J_{1}=1$.  The meaning of the $E_{\mathrm{max}}$ points 
shown is described in the text.}
\label{E_spin1}
\end{figure*}

Figure~\ref{E_spin1}
shows the extrapolated CCM results for the gs energy per spin, $E/N$, as a function 
of $J_{2}$ for various values of $\Delta$, using both the $z$-aligned and planar $x$-aligned 
model states.  For each value of $\Delta$ two curves are shown, one (for smaller 
values of $J_{2}$) using the N\'{e}el state, and the other (for larger values of 
$J_{2}$) using the stripe state as CCM model state.  As has been discussed in 
detail elsewhere~\cite{Bi:1998_b,Ze:1998,Fa:2004}, 
the coupled sets of LSUB$n$ equations (\ref{ket_coeff}) have natural termination points 
(at least for values $n > 2$) for some critical value of a control parameter 
(here the anisotropy, $\Delta$), beyond which no real solutions to the equations exist.  
Thus, for each set of calculations based on one of the four CCM model states used, 
the $E_{\mathrm{max}}$ points shown in figure \ref{E_spin1} are either those natural termination points 
described above for the highest (SUB8--8) level of approximation we have implemented, or the 
points where the gs energy becomes a maximum should the latter occur first (i.e., as one 
approaches the termination point).  The advantage of this usage of the $E_{\mathrm{max}}$ 
points is that we do not then display gs energy data in any appreciable regimes 
where SUB$n$--$n$ calculations with very large values of $n$ (higher than can feasibly 
be implemented) would not have solutions, because of having terminated already.

All of the curves such as those shown in figure~\ref{E_spin1} illustrate very clearly that the 
corresponding pairs of gs energy curves (for the same values of $\Delta$) for the N\'{e}el 
and stripe phases cross one another, for both the $z$-aligned (figure~\ref{E_Zaligned_spin1} 
for all values $\Delta > 1$) and the $x$-aligned (figure~\ref{E_planar_spin1} for all values 
$0 \leq \Delta < 1$) cases.  The crossings occur with a clear discontinuity in slope, which 
is completely characteristic of a first-order phase transition, exactly as observed in 
the classical (i.e., $s \rightarrow \infty$) case.  Unlike in the $s = \frac{1}{2}$ version 
of this model that we studied earlier~\cite{Bi:2008_spinHalf_J1J2anisotrtopy}, there is no 
indication at all in the present $s=1$ case of any intermediate paramagnetic phase emerging 
for any values of the parameters $J_2$ and $\Delta$.  Furthermore, the direct first-order 
phase transition, so indicated by our results for the gs energy, 
between the quasiclassical N\'{e}el-ordered and collinear stripe-ordered phases,  
in both the $z$-aligned and planar $x$-aligned cases, occurs for all values of $\Delta \geq 0$ 
very close to the classical phase boundary $J_{2}^{c}=\frac{1}{2}$, the point of 
maximum (classical) frustration.  

\begin{figure*}[tbp]
\begin{center}
\mbox{
  \subfigure[$z$-aligned states]{\label{M_Zaligned_spin1}\scalebox{0.3}{\epsfig{file=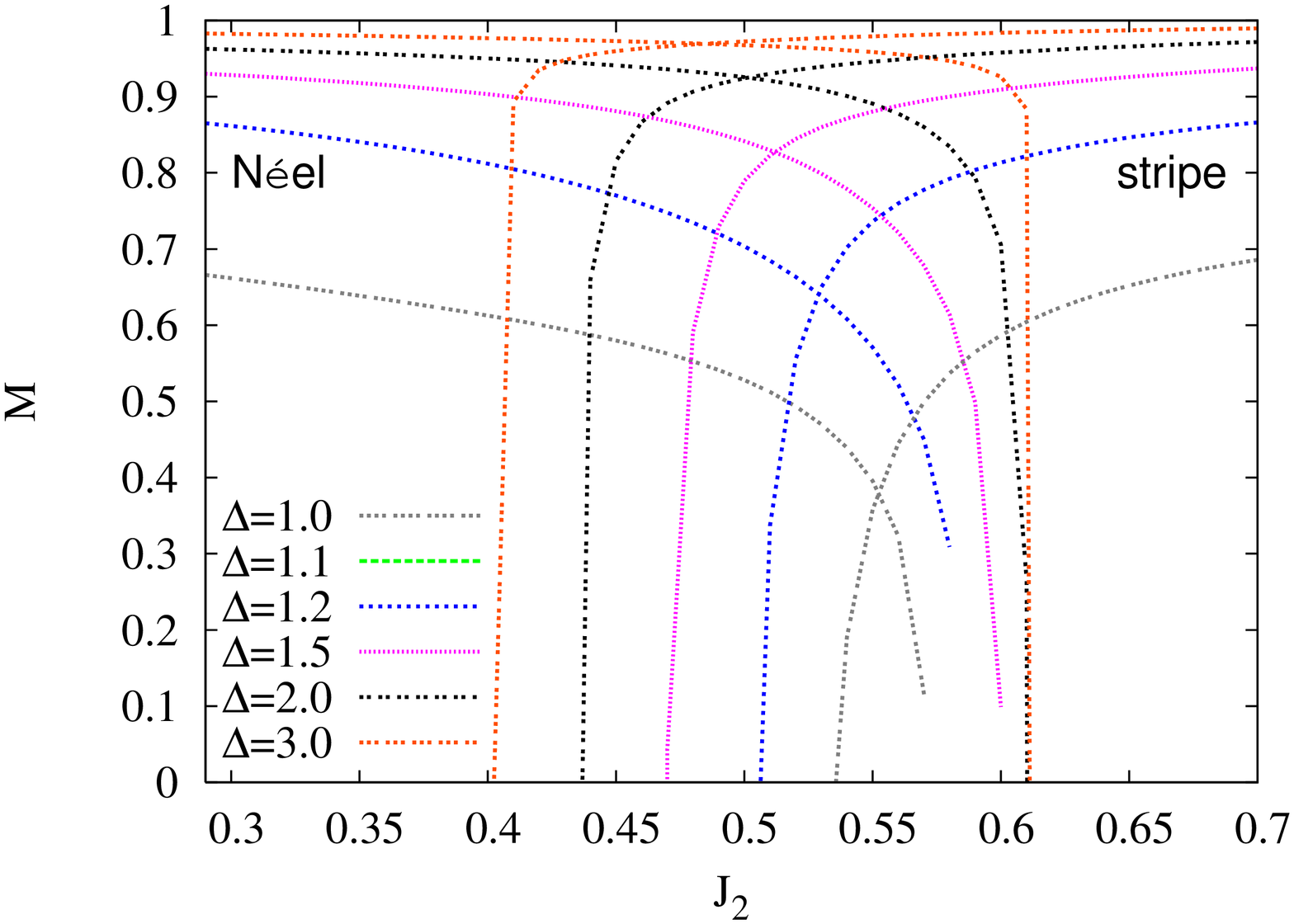,angle=270}}}
  \subfigure[planar $x$-aligned states]{\label{M_planar_spin1}\scalebox{0.3}{\epsfig{file=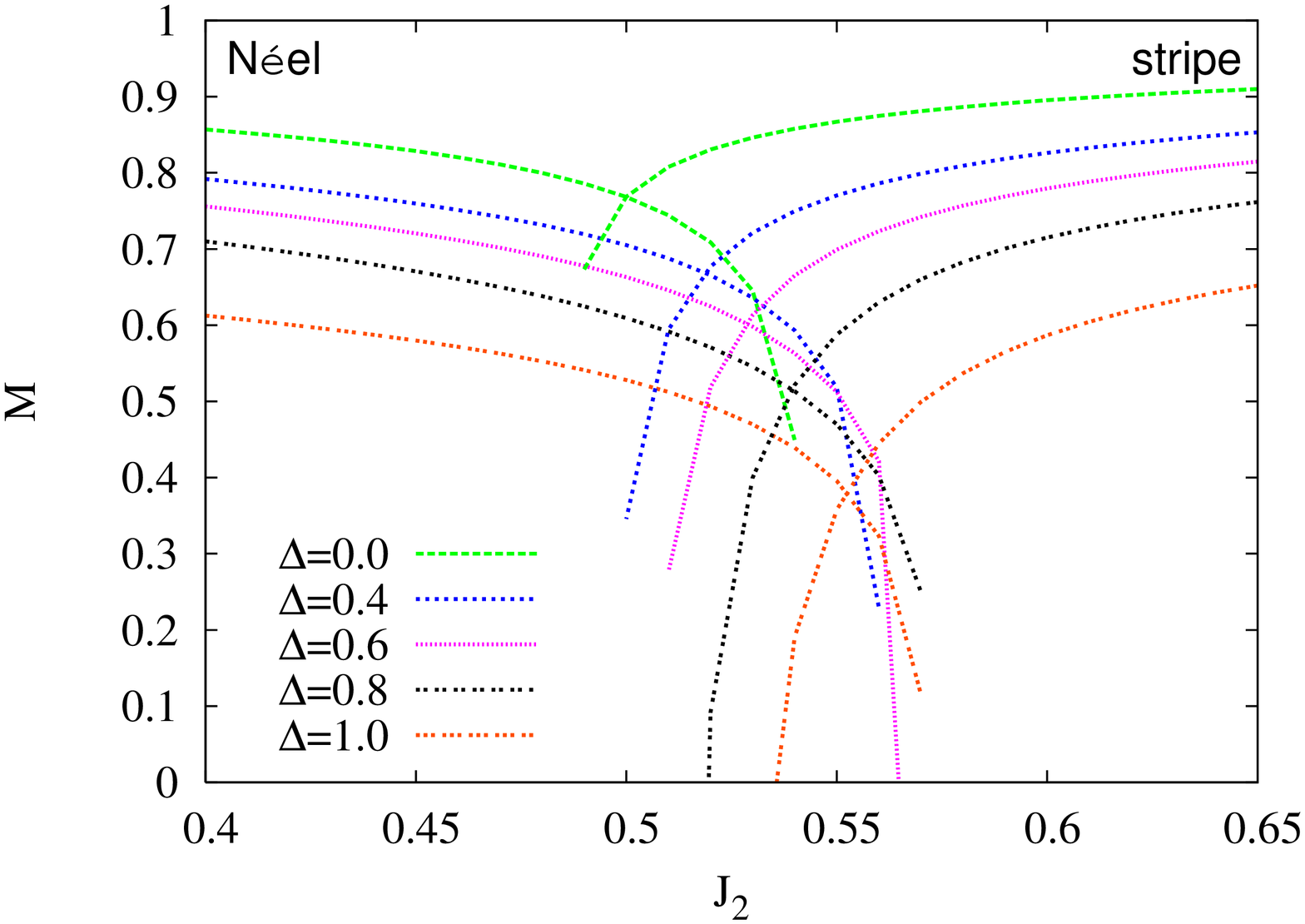,angle=270}}}
}
\caption{(Colour online) Extrapolated CCM SUB$n$--$n$ results using the $z$-aligned and 
planar $x$-aligned states for the gs staggered magnetization, $M$, for the N\'{e}el 
and stripe phases of the $s=1$ $J_{1}^{XXZ}$--$J_{2}^{XXZ}$ model.  The SUB$n$--$n$ 
results are extrapolated to the limit $n \rightarrow \infty$ using the sets $n=\{2,4,6,8\}$ 
for both the $z$-aligned state and the planar $x$-aligned states.  
The NN exchange coupling $J_{1}=1$.}
\label{M_spin1}
\end{center}
\end{figure*}

We show in figure~\ref{M_spin1}
corresponding indicative sets of CCM results, based on the same four model states, for the 
gs order parameter (viz., the staggered magnetization), to those shown in figure \ref{E_spin1} for 
the gs energy.  The staggered magnetization data completely reinforce the phase structure 
of the model as deduced above from the gs energy data.  

Thus, let us now denote by $M_{c}$ the quantum phase transition point 
deduced from curves such as those shown in figure \ref{M_spin1}, 
where $M_{c}$ is generically defined to be either (a) the point where corresponding pairs 
of CCM staggered magnetization curves (for the same value of $\Delta$), based on 
the N\'{e}el and stripe model states, intersect one another if they do so at a physical 
value $M \geq 0$; or (b) if they do not so intersect at a value $M \geq 0$, the 
two points where the corresponding values of the staggered magnetization go to zero.  
Clearly, in this generic scenario, case (a) corresponds to a direct phase 
transition between the N\'{e}el and 
stripe phases, which will generally be first-order if the intersection point has a value 
$M \neq 0$ (and, only exceptionally, second-order, if the crossing occurs exactly at $M = 0$).  
On the other hand, case (b) corresponds to the situation where the points where the LRO 
vanishes for both quasiclassical (i.e., N\'{e}el-ordered and stripe-ordered) phases are 
indicative of a phase transition from each of these phases 
to some intermediate magnetically-disordered phase.  A detailed discussion 
of this order parameter criterion for a phase transition and its relation to the 
stricter energy crossing criterion has been given elsewhere~\cite{Schm:2006}.  

It is clear from figures~\ref{M_Zaligned_spin1} and~\ref{M_planar_spin1} that case (b) above 
never occurs for the present spin-1 model for any values of the anisotropy parameter $\Delta$ 
or for any values of the NNN exchange coupling $J_2$, unlike in the $s = \frac{1}{2}$ version 
of this model that we studied earlier~\cite{Bi:2008_spinHalf_J1J2anisotrtopy}. 

\begin{figure}[tbp]
\begin{center}
\epsfig{file=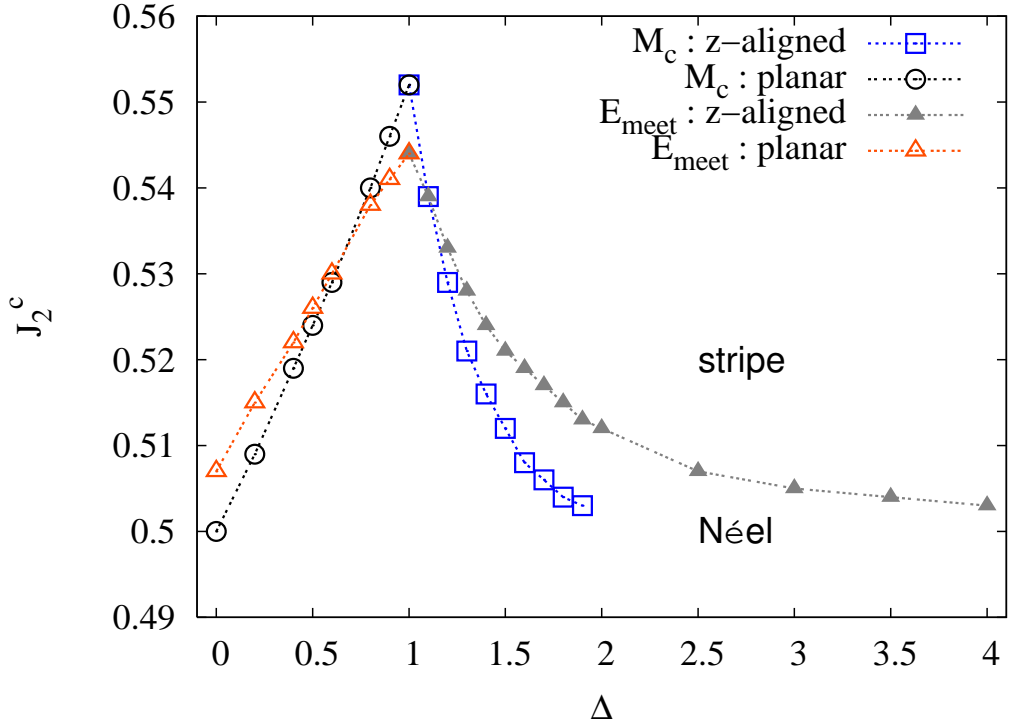,width=10cm,angle=270}
\caption{(Colour online) Extrapolated CCM SUB$n$--$n$ results using the $z$-aligned and planar 
$x$-aligned states for the ground-state phase diagram of the $s=1$ $J_{1}^{XXZ}$--$J_{2}^{XXZ}$ 
anisotropic Heisenberg model on the square lattice, for the NN exchange coupling $J_{1}=1$.  
The SUB$n$--$n$ results for the energy per spin and the staggered magnetization are 
extrapolated to the limit $n \rightarrow \infty$ using the sets $n = \{2,4,6,8\}$ for both 
the z-aligned and planar $x$-aligned model states.  $M_{c} \equiv $ magnetization critical point, 
defined in the text.  $E_{\mathrm{meet}}$ denotes the crossing point of the CCM energy curves 
for the same value of $\Delta$ based on the N\'{e}el-ordered and collinear stripe-ordered 
model states.}
\label{phase_spin1}
\end{center}
\end{figure}

By putting together data of the sort shown in figures \ref{E_spin1} and \ref{M_spin1} we can 
now deduce the gs phase diagram of our system from our CCM calculations based on the four model 
states with quasiclassical antiferromagnetic LRO that we have employed.  Figure~\ref{phase_spin1}
shows the zero-temperature gs phase diagram of the 2D $s=1$ $J_{1}^{XXZ}$--$J_{2}^{XXZ}$ model 
on the square lattice for the $z$-aligned and planar $x$-aligned states, as obtained from our 
extrapolated results for both the gs energy and the gs order parameter.  The 
completely independent results from both the energy criterion 
and the order parameter criterion for the phase transition give extremely similar positions for 
the phase boundary, as one can observe from figure~\ref{phase_spin1}.  Note that the results from 
using the order parameter criterion become increasingly inaccurate for large values of $\Delta$, 
and this is why we show them in figure~\ref{phase_spin1} only out to $\Delta \lesssim 2$.  
The reason for this is simple.  Thus, as $\Delta \rightarrow \infty$, the order parameters $M \rightarrow 1$ 
for both the N\'{e}el-ordered and collinear stripe-ordered phases, and it becomes 
increasingly difficult to determine the point where they cross, since the angle of their crossing 
becomes vanishingly small.  This effect can clearly be seen in figure~\ref{M_Zaligned_spin1}, 
where it has clearly become acute even for values of $\Delta$ as small as about 2.  On the other hand, 
the energy criterion correspondingly becomes {\it more\/} accurate as $\Delta \rightarrow \infty$, 
as one may observe from figure~\ref{E_Zaligned_spin1}.  Thus, figure~\ref{phase_spin1} clearly shows that 
the phase boundary approaches the classical line $J_{2}^{c} = 0.5$ as $\Delta \rightarrow \infty$, as
expected in this Ising-like limit.

Our results certainly provide very clear and consistent evidence that there exists no intermediate 
phase.  Thus, the curves for the order parameters of the N\'{e}el and stripe phases 
always meet at a finite value and the corresponding curves for the gs energies of the two 
phases intersect with a discontinuity in slope, for both the $z$-aligned and planar 
$x$-aligned states, for all values of the anisotropy parameter $\Delta$.  All of the evidence 
clearly points towards a first-order phase transition between the two phases.  

We note also that the $z$-aligned and $xy$-planar-aligned phases meet precisely at the isotropic point $\Delta = 1$, 
just as in the classical case, and exactly as expected.  However, this does provide a consistency 
check on our independent numerical calculations for the two phases.  The case $\Delta = 1$ obviously 
reproduces the usual (isotropic) $J_{1}$--$J_{2}$ model.  Thus, at $\Delta = 1$, we find 
$J_{2}^{c} = 0.55 \pm 0.01$ which, very encouragingly, is the same value we found~\cite{Bi:2008} 
for the $s=1$ $J_{1}$--$J_{1}'$--$J_{2}$ model in the spatially isotropic limiting case when 
$J_{1}'/J_{1} = 1$.  We also note that in the present spin-1 quantum model, the isotropic point 
$\Delta = 1$ is precisely the point at which the boundary between the two quasiclassical phases deviates 
most from its classical position at $J_{2}^{c} = \frac{1}{2}$ for all values of $\Delta \geq 0$.  
Our calculations also indicate that at the isotropic $XY$ point of the model (i.e., where 
$\Delta = 0$) the phase boundary is at $J_{2}^{c} = 0.50 \pm 0.01$.

\section{Discussion}
Our results have clearly shown in detail how the quantum fluctuations present in 
the spin-1 $J_1$--$J_2$ model on the infinite square lattice are diminished by 
varying the spin anisotropy parameter $\Delta$ away from the Heisenberg isotropic 
point $\Delta = 1$ in either direction.  This is precisely as was observed 
previously~\cite{Bi:2008_spinHalf_J1J2anisotrtopy} for the spin-$\frac{1}{2}$ version 
of the same model, and as was to be expected.  However, unlike what would be 
predicted by lowest-order (or linear) spin-wave theory (LSWT)~\cite{Ch:1988}, for 
example, we can now conclude with confidence from our results that no such 
intermediate disordered phase as the one that we observed in the spin-$\frac{1}{2}$ 
version of this model between the two quantum triple points at 
($\Delta ^{c}$ = $-0.10 \pm 0.15$, $J_{2}^{c}/J_{1}$ = $0.505 \pm 0.015$) and 
($\Delta ^{c}$ = $2.05 \pm 0.15$, $J_{2}^{c}/J_{1}$ = $0.530 \pm 0.015$), 
exists for the spin-1 version, for any values of the parameters $J_2/J_1$ 
and~$\Delta$. 

In the context of a spin-wave theory (SWT) treatment of the isotropic $J_1$--$J_2$ model on the 
square lattice, LSWT predicts that quantum fluctuations can destabilize the 
classical GS with LRO, even at large values of the spin quantum number $s$, for 
values of the frustration parameter $J_{2}/J_{1}$ around 0.5.  For the 
spin-$\frac{1}{2}$ case the range of values, $\alpha^{c_1} < J_{2}/J_{1} < \alpha^{c_2}$, 
for which a magnetically-disordered phase thereby occurs 
is predicted by LSWT to be given by $\alpha^{c_1} \approx 0.38$ and 
$\alpha^{c_2} \approx 0.52$.  These values may be  
compared to our own predictions~\cite{Bi:2008_spinHalf_J1J2anisotrtopy} of 
$\alpha^{c_1} = 0.44 \pm 0.01$ and $\alpha^{c_2} = 0.59 \pm 0.01$.  For the spin-1 
case LSWT predicts a narrower, but still non-vanishing, strip of disordered 
intermediate phase in a range with 
$\alpha^{c_1} \approx 0.47$ and $\alpha^{c_2} \approx 0.501$, whereas we predict 
with confidence that the disordered phase simply does not exist as a GS in this case.  

The discrepancy between our results and those of LSWT for the spin-1 case are 
undoubtedly due to the shortcomings of LSWT.  Thus, while LSWT can work reasonably 
well in the absence of frustration (e.g., for the isotropic $J_1$--$J_2$ model 
here when $J_2 = 0$, that represents the Heisenberg model with only NN interactions), 
in the presence of frustration it consistently overestimates the effects of quantum 
fluctuations.  This effect worsens as the frustration (here measured by the ratio 
$J_{2}/J_{1}$) increases.

Thus, Igarashi~\cite{Ig:1993} has shown explicitly for the $J_1$--$J_2$ model by 
going to higher orders in SWT (i.e., by calculating higher-order terms in the 
$1/s$ power expansion), that while the series seems to converge for values 
$J_{2}/J_{1} \lesssim 0.35$, the second-order corrections grow so large for values 
$J_{2}/J_{1} \gtrsim 0.4$ that no prediction based on LSWT, or even on higher-order 
SWT, in this region (e.g., about the appearance of an intermediate 
magnetically-disordered phase near $J_{2}/J_{1} \approx 0.5$) should be relied upon.  
Furthermore, he showed that the effects of the higher-order correction terms to 
LSWT make the N\'{e}el-ordered state more stable than predicted by LSWT.

Relatively little attention has been paid by other authors to the (pure, isotropic) 
$J_1$--$J_2$ model at higher values of the spin quantum number, $s > \frac{1}{2}$.  
We note, however, that Cai {\it et al.}~\cite{CCKW:2007} have also recently 
postulated the possible existence of an intermediate phase between the quasiclassical 
N\'{e}el-ordered and collinear stripe-ordered phases for the spin-1 model.  More 
specifically, they hypothesize an intermediate valence-bond solid (VBS) ground state (GS) 
for the spin-1 isotropic $J_1$--$J_2$ model at or near the point of maximal classical 
frustration where $J_{2}/J_{1} = 0.5$.  Their evidence is indirect and is based on a 
trial variational state of VBS type, which is an {\it exact\/} GS of a related spin-1 
model Hamiltonian, and on a pseudopotential approach to extend it to the actual spin-1 
$J_1$--$J_2$ model.  They express the dual hopes that this trial state might capture 
the main character of the disordered phase that they thereby predict for the fully 
frustrated case, and that accurate numerical methods, such as those considered here, 
might verify the existence of this postulated intermediate phase.  Such variational 
analyses, based on physically motivated trial states, are always of interest, but have 
a very chequered history of success in the field of highly correlated spin- and 
electron-lattice systems.  In the present case we stress again that our own detailed 
numerical analysis provides no evidence at all for the existence of such an 
intermediate magnetically-disordered VBS phase as postulated by 
Cai {\it et al.}~\cite{CCKW:2007}.

In the same context, we note too that in earlier work Read and Sachdev~\cite{RS:1991} 
have applied a large-$N$ expansion technique based on symplectic Sp($N$) symmetry to the 
isotropic $J_1$--$J_2$ model.  They found that the method, which can 
itself be regarded as akin to a $1/s$ expansion, predicts an intermediate 
phase (with VBS order) for smaller values of $s$, but that this 
phase disappears for larger values of $s$ where they predict 
instead a first-order transition between the N\'{e}el and stripe phases.  All 
of these qualitative results for the pure $J_1$--$J_2$ model are in accord with 
our quantitative predictions.

We note that the results presented here for the spin-anisotropic spin-1 
$J_1^{XXZ}$--$J_2^{XXZ}$ model are also fully consistent with our own 
previous results~\cite{Bi:2008} for the spatially-anisotropic spin-1 
$J_1$--$J_{1}'$--$J_{2}$ model discussed in section~\ref{intro} above, for 
which we also found no evidence for an intermediate disordered phase between the 
quasiclassical N\'{e}el and collinear stripe phases with LRO.  However, whereas 
for the spin-1 $J_1$--$J_{1}'$--$J_2$ model we found strong evidence for a 
quantum tricritical point at ($J_{1}'/J_1 \approx 0.66$, $J_2/J_1 \approx 0.35$) 
where a line of second-order phase transitions between the N\'{e}el-ordered 
and the collinear stripe-ordered states (for $J_{1}'/J_1 \lesssim 0.66$) meets 
a line of first-ordered phase transitions between the same two states (for 
$J_{1}'/J_1 \gtrsim 0.66$), we find for the present spin-1 $J_{1}^{XXZ}$--$J_{2}^{XXZ}$ 
model that the phase transition between these two states is first-order for all 
values $\Delta \geq 0$.  Clearly, these two sets of results are in complete 
agreement with one another at their common point of overlap, when $J_{1}' = J_{1}$ and 
$\Delta = 1$.

At the $XY$ isotropic point ($\Delta = 0$) of the present spin-1 
$J_{1}^{XXZ}$--$J_{2}^{XXZ}$ model we predict that the phase boundary occurs at a 
value $J_2^c(0) = 0.50 \pm 0.01$.  It is interesting to note that our previous 
results for the spin-$\frac{1}{2}$ version of the model~\cite{Bi:2008_spinHalf_J1J2anisotrtopy} 
showed a quantum triple point (QTP) at ($\Delta^c = -0.10 \pm 0.015$, $J_2^c = 0.505 \pm 0.015$).  
Clearly our results for this spin-$\frac{1}{2}$ case are consistent with this lower QTP 
occurring exactly at the $XY$ isotropic point ($\Delta = 0$) and also at the point of 
maximum classical frustration, $J_2 = \frac{1}{2}$.  Similarly, in the present spin-1 
case our results are consistent with the phase boundary at the $XY$ isotropic point also 
occurring at the point $J_2 = \frac{1}{2}$.  It would seem likely, therefore, that for 
both the cases of spin-$\frac{1}{2}$ and spin-1 particles the corresponding quantum 
$J_1^{XX}$--$J_2^{XX}$ model has a special behaviour at the point $J_2/J_1 = \frac{1}{2}$ 
where the classical frustration is greatest.  Our results indicate that a more detailed 
investigation of this case might, therefore, be worth undertaking for general values 
of the spin quantum number $s$.

Although there is very little other accurate numerical work for the present model 
against which to make comparisons, there have been several previous detailed 
comparisons, for example, of CCM results with those from the exact diagonalization (ED) 
of finite spin-lattices for some particular models.  One such 
example~\cite{Kr:2000} is the spin-$\frac{1}{2}$ $J$--$J'$ (or zigzag) 
model on the square lattice which contains two kinds of NN isotropic Heisenberg 
interactions, of strength $J$ and $J'$ respectively, such that each square plaquette 
contains three $J$-bonds and one $J'$-bond, with the $J'$-bonds arranged in a regular zigzag 
fashion such that every lattice site on the square lattice is joined to only one 
$J'$-bond.  An alternative but equivalent description of the model is that it 
interpolates between a honeycomb and a square lattice, such that the $J$-bonds join 
NN lattice sites on the honeycomb lattice, and the $J'$-bonds join sites across only 
one of the main diagonals of each hexagon, such that when $J=J'$ the model is equivalent 
to the NN isotropic Heisenberg model on the square lattice.

ED calculations were performed for the above model~\cite{Kr:2000} for lattices with up to 
$N=32$ sites.  In general terms it was found that the CCM results for the model at 
attainable levels of implementation (viz., using the LSUB$n$ approximation with $n \leq 8$ 
agree well with the extrapolated ($N \rightarrow \infty$) ED data.  The CCM is particularly 
good, however, at describing both the dimerized and the helical gs phases that this 
system can support.  For the latter phase the ED results lie appreciably above those 
from the CCM.  This is because the energies for the small lattices able to be considered 
do not fit well to the known theoretical finite-size scaling law in this regime.  It 
is no surprise that finite-size effects for systems with an incommensurate helical spin 
structure are larger than for systems with N\'{e}el order or that are ordered with 
dimerized spin pairs.

Similar conclusions were also drawn for comparions of CCM and ED results for extensions 
of the above spin-$\frac{1}{2}$ $J$--$J'$ model to both (a) the anisotropic 
$J_{XXZ}$--$J'_{XXZ}$ model~\cite{Da:2004} where both bonds contain an Ising anisotropy 
of precisely the sort considered in the present paper; and (b) the case where the spin 
quantum number $s > \frac{1}{2}$~\cite{Da:2005}.  For the latter case of the spin-1 
$J$--$J'$ model, calculations were performed using both the CCM in the SUB$n$--$n$ scheme 
with $n \leq 6$ and the ED technique on lattices of sizes $N \leq 20$.  Again, the 
resulting finite-size ED extrapolations remained quite poor, and only allowed some 
qualitative conclusions to be drawn, whereas results from the CCM were seen to be much 
more robust and more reliable.  In no case, however, did the CCM and ED results 
conflict with each other.

Another model where ED and CCM results have been compared is the pure (isotropic) 
spin-$\frac{1}{2}$ $J_1$--$J_2$ model on the square lattice~\cite{DDZSKR:2008}.  
Again, the ED results (with $N \leq 32$) were found to provide a good qualitative 
check of the CCM data for LSUB$n$ calculations performed with $n \leq 8$.
Finally, for the spin-$\frac{1}{2}$ version of the 
present anisotropic $J_{1}^{XXZ}$--$J_{2}^{XXZ}$ model, 
we~\cite{DRSBL:2008} have also compared the CCM results with those from ED calculations on 
finite-sized lattices of size $N =36 = 6 \times 6$ sites (with periodic boundary conditions 
imposed).  In this case too the ED data are best used to complement the CCM results.  
On the basis of all the above evidence we expect that the same will hold true for the 
spin-1 version of the model studied here.  Since the number of basis states increases 
roughly as $3^N$ for the spin-1 case, by comparison with $2^N$ for the spin-$\frac{1}{2}$ 
case, ED calculations for the present model would be limited to lattices of sizes $N = 16$ 
and $N = 20$.  The next biggest lattice that preserves the full lattice symmetry has 
$N = 26$ sites in this case, and an ED calculation of this size for the spin-1 model is 
probably beyond the limits of presently available computing power.  With only such 
limited data the ED finite-size extraploation would again be bound to remain poor, as 
seen in the previous work cited above, and we fully expect that the CCM results 
would again prevail even if ED results were available for the present model.

Finally, we note that our analysis and conclusions have relied heavily on 
two of the unique strengths of the CCM, namely its ability
to deal with highly frustrated systems as easily as unfrustrated
ones, and its use from the outset of infinite lattices.
These, in turn, lead to its ability to yield accurate predictions for the locations of phase
boundaries.  Our own results for the gs energy and staggered magnetization
provide a set of independent checks that lead us
to believe that we now have a self-consistent and coherent
description of these challenging anisotropic and frustrated $J_{1}^{XXZ}$--$J_{2}^{XXZ}$ systems for both
the spin-$\frac{1}{2}$ and spin-1 cases.

\section*{Acknowledgements}
We thank the University of Minnesota Supercomputing 
Institute for Digital Simulation and Advanced Computation  
for the grant of supercomputing facilities in conducting this research.  Two of 
us (RD and JR) are grateful to the DFG for support (through project Ri615/16-1).  
We also thank Zi Cai~\cite{CCKW:2007}, Zhong-Yi Lu~\cite{MLX:2008}, 
Subir Sachdev~\cite{RS:1991} and Qimiao Si~\cite{SA:2008} for bringing 
their respectively cited papers to our attention.

\end{document}